\documentclass[doublecol]{epl2}
\usepackage{graphicx}
\usepackage{amsmath}
\usepackage{revsymb}
\usepackage{bm}
\newcommand{\s}{\sum\limits}

\newcommand{\pa}{\partial}

\newcommand{\be}{\begin{equation}}
\newcommand{\e}{\end{equation}}
\newcommand{\beml}{\begin{subequations}}
\newcommand{\eml}{\end{subequations}}
\newcommand{\beq}{\begin{eqnarray}}
\newcommand{\eq}{\end{eqnarray}}
\newcommand{\ba}{\begin{array}}
\newcommand{\ea}{\end{array}}
\newcommand{\lt}{\left}
\newcommand{\rt}{\right}
\newcommand{\n}{\nonumber}
\newcommand{\la}{\langle}
\newcommand{\ra}{\rangle}
\newcommand{\ep}{\varepsilon}

\newcommand{\bb}{\boldsymbol}

\DeclareMathOperator{\diag}{diag}

\begin{document}

\title{Impurity-assisted tunneling in graphene}
\author{M. Titov}
\institute{Department of Physics, Konstanz University, 
D--78457 Konstanz, Germany}
\date{January 2007}

\abstract{
The electric conductance of a strip of undoped graphene increases
in the presence of a disorder potential, which is smooth on atomic scales.
The phenomenon is attributed to impurity-assisted resonant tunneling 
of massless Dirac fermions. Employing the transfer matrix approach 
we demonstrate the resonant character of the conductivity enhancement
in the presence of a single impurity. We also calculate the two-terminal 
conductivity for the model with one-dimensional fluctuations of 
disorder potential by a mapping onto a problem of Anderson localization. 
}

\maketitle

A monoatomic layer of graphite, or graphene, has been 
recently proven to exist in nature \cite{Nov04,Nov05,Zha05a,Zha05b}. 
Low-energy excitations in graphene are described by 
the "relativistic" massless Dirac equation, 
which gives us theoretical insight into exotic 
transport properties observed in this material. Undoped graphene is 
a gapless semiconductor,
or semi-metal, with vanishing density of states at the Fermi level. 
One of the first experiments \cite{Nov05} shows 
that the conductivity of graphene at low temperatures takes on a nearly universal value of the order 
of $4 e^2/h$ and increases if a doping potential of any polarity is applied. 
Systematic dependence of the minimal conductivity on the sample size 
has been observed recently in Ref.~\cite{Lau07}.

The peculiar band-structure of the two-dimensional carbon,
which mainly explains many recent experimental observations, 
has already been calculated in 1947 by Wallace \cite{Wal47}.
Nevertheless, the universal value of the minimal conductivity 
is not entirely understood. Many recent theoretical studies  
\cite{Per06,Zie06,Cse06,Ost06,Ale06,Alt06} address the problem of  
the finite conductivity of the undoped graphene by employing the Kubo formula. 
Other works \cite{Kat05,Two06,Ryu06} show that the conductance $G$ 
of a ballistic graphene sample (of the width $W$ much larger 
than the length $L$) scales as 
$G= \sigma W/L$ with the coefficient $\sigma=4 e^2/\pi h$. 
This value of $\sigma$ coincides with the prediction made for the $dc$ conductivity 
of disordered graphene \cite{Lud94,Gor02,Per06,Zie06,Cse06,Ost06} 
and agrees with experiments done on small samples \cite{Lau07}.

In this work we develop an extension of the transfer matrix formalism 
of Refs.~\cite{Two06,Tit06,Che06} in order to include 
the effects of disorder. Our main result is the enhancement 
of the zero temperature conductance at low doping by an impurity potential, 
which is smooth on atomic scales. (Such potential corresponds to 
a diagonal term in the Dirac Hamiltonian \cite{Sho98}).
Our results agree with recent numerical studies \cite{Ver06,Ryc06}
and with a related work \cite{Nom06}, where the conductivity 
enhancement by smooth disorder with infinite correlation range
was predicted.

\begin{figure}[tb]
\centerline{\includegraphics[width=0.9\linewidth]{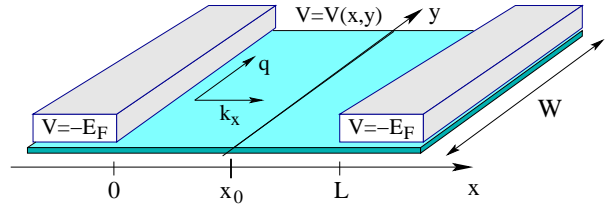}}
\caption{\label{fig:setup} 
A ribbon of undoped graphene is contacted by two metallic leads.
The charge carriers tunnel from one lead to another via multiple 
tunneling states formed in the graphene strip. 
For clean sample with $L\ll W$ the conductance $G$ scales as 
$G=g_0 W/\pi L$, where $g_0= 4e^2/h$ is the conductance 
quantum in graphene and $L$ is the length of the strip. 
An impurity placed inside the strip enhances the conductance
in a vicinity of the Dirac point $\ep \ll \hbar v/L$,  
provided the impurity strength is close to one of the resonant values. 
}
\end{figure}

We analytically calculate the two-terminal conductivity 
of a graphene sheet in a model with one-dimensional fluctuations 
of the disorder potential taking advantage of 
a mapping onto a problem of Anderson localization. 
In this model the conductivity is found to increase as the square 
root of the system size. We find that quantum interference effects 
are responsible for the leading contribution to 
the conductivity of graphene near the Dirac point.


We start by considering the effects of a single impurity 
in the setup depicted in Fig.~\ref{fig:setup}. 
At low doping the conductance is determined by quasiparticle tunneling, 
which is independent on the boundary conditions in $y$-direction if $L\ll W$. 
(For illustrations we choose periodic boundary conditions with $W/L=7$). 
We find that a single impurity placed in an ideal sheet of undoped 
graphene modifies the tunneling states and leads to the conductivity 
enhancement provided the impurity strength is close to one of the 
multiple resonant values. Away from the Dirac point 
the presence of an impurity causes a suppression of the conductance. 

In this study we restrict ourselves to the single-valley 
Dirac equation for graphene,
\be
\label{H}
-i \hbar v\, \bb{\sigma}\cdot \bb{\nabla}\Psi  + V \Psi = \ep\Psi, 
\e
where $\Psi$ is a spinor of wave amplitudes for two non-equivalent 
sites of the honeycomb lattice. The Fermi-energy $\ep$ and the impurity 
potential $V(x,y)$ in graphene sample ($0<x<L$) are considered to be much smaller 
than the Fermi-energy $E_F$ in the ideal metallic leads ($x<0$ and $x>L$). 
For zero doping the conductance is determined by the states at 
the Dirac point, $\ep=0$. Transport properties at finite energies 
determine the conductance of doped graphene. 

The Dirac equation in the leads has a trivial solution
$\Psi \sim \exp(\pm i {\bb{k}} {\bb{r}})$ with the wave vector
${\bb{k}}=(k_x , q)$ for the energy $\ep=\hbar v \sqrt{k_x^2+q^2}-E_F$.    
In order to make our notations more compact we let $\hbar v=1$ in the
rest of the paper. The units are reinstated in the final results and 
in the figures. For definiteness we choose periodic boundary 
conditions in $y$ direction, hence the transversal momentum $q$ is quantized as
$q_n=2\pi n/ W$, with $n=0,\pm 1, \pm 2, \dots \pm M$. The value of $M$ is determined by
the Fermi energy $E_F$ in the leads, $M={\rm Int}\;[W/\lambda_F]$,
where $\lambda_F=2 \pi/E_F$, and the number of propagating channels 
is given by $N=2 M+1$. 

For $\ep\ll E_F$ the conductance is dominated by modes with 
a small transversal momentum $q_n\ll k_F$. The corresponding   
scattering state for a quasiparticle injected from the left lead 
is given by
\beq
\n
\Psi_n^{(L)} &=& \chi e^{i{\bb{k}}_n {\bb{r}}} + \phi \s_m r_{nm} 
e^{-i{\bb{k}}_m {\bb{r}}},\quad x<0,\\
\Psi_n^{(R)} &=& \chi \s_m t_{nm} e^{i{\bb{k}}_m {\bb{r}}},\quad x>L,
\label{Psi}
\eq
where ${\bb{k}}_n = (\sqrt{k_F^2-q_n^2},q_n) \approx (k_F, q_n) $, and 
\be
\chi=\frac{1}{\sqrt{2}}\lt(\ba{c}1\\ 1\ea\rt),\qquad 
\phi=\frac{1}{\sqrt{2}}\lt(\ba{c}1\\ -1 \ea\rt).
\e
The conductance of the graphene strip is expressed through 
the transmission amplitudes $t_{nm}$ in Eq.~(\ref{Psi}) 
by the Landauer formula,
\be
\label{G}
G=g_0\s_{n,m}|t_{nm}|^2, \qquad g_0=4 e^2/ h,
\e
where the summation extends from $-M$ to $M$. 
The factor of $4$ in the conductance quantum 
is due to the additional spin and valley degeneracies.

In order to find $t_{nm}$ we have to solve the scattering problem.
The solution becomes more transparent 
if one takes advantage of the unitary rotation in the isospin space 
${\cal L}=(\sigma_x+\sigma_z)/\sqrt{2}$, 
which transforms the spinors $\chi$ and $\phi$ into $(1,0)$ and $(0,1)$, correspondingly.
We combine such a rotation with the Fourier transform in the transversal direction
and arrange the spinors
\be
\psi_n(x)={\cal L} \,\frac{1}{W}\int_0^W dy\; e^{i q_n y}\Psi(x,y)
\e
in the vector ${\bb{\psi}}(x)$ of length $2N$.
Then, the evolution of ${\bb{\psi}}(x)$ inside the graphene sample 
can be written as
${\bb{\psi}}(x)= {\cal T}_x {\bb{\psi}}(0)$,
where the $2N\times 2N$ transfer matrix ${\cal T}_x$ fulfills the 
flux conservation law 
${\cal T}_x^\dagger \sigma_z {\cal T}_x =\sigma_z$. 
In the chosen basis the transfer matrix of the whole sample is straightforwardly 
related to the matrices of transmission and reflection amplitudes,  
\be
\label{param}
{\cal T}\equiv {\cal T}_L = 
\lt(\ba{cc}\hat{t}^{\dagger -1} & \hat{r}^{\prime} \hat{t}^{\prime -1} \\ 
- \hat{t}^{\prime -1} \hat{r} & \hat{t}^{\prime -1} \ea\rt),
\e
defined in the channel space.

\begin{figure}[tb]
\centerline{\includegraphics[width=0.9\linewidth]{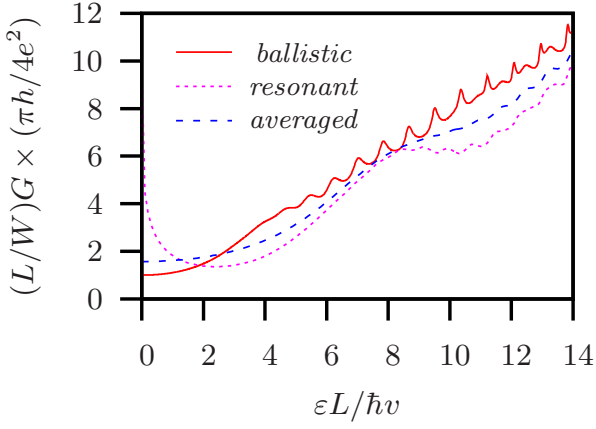}}
\caption{\label{fig:doping} 
The conductance of the graphene strip, $W/L=7$, with a 
potential interface (\ref{const}) in the middle
of the sample. The curves are calculated from 
Eqs.~(\ref{G},\ref{param},\ref{T0a}). 
The solid line shows the conductance of the {\em ballistic} sample, 
$\alpha=0$. The dashed line corresponds to the case of 
a {\em resonant} impurity strength, $\alpha = \pi/2+\pi n$. The 
{\em averaged} conductance for the stochastic model with $\alpha$, 
which is uniformly distributed 
in the interval $(0,2\pi)$, is plotted with the dotted line.
}
\end{figure}

The equation for the transfer matrix follows from the Dirac equation (\ref{H}),
\be
\label{T}
\frac{\pa {\cal T}_x}{\pa x}=
\lt(\sigma_x\otimes\hat{q} + i\sigma_z\otimes(\ep\hat{\openone}-\hat{V}(x))\rt){\cal T}_x,
\e
where $\hat{q}$ is a diagonal matrix with entries $q_n$ 
and $\openone$ is the unit matrix in the channel space. 
The elements of $\hat{V}(x)$ are given by
\be
\label{FT}
V_{nm}(x)=\frac{1}{W}\int_0^W dy\;e^{i(q_n-q_m)y}V(x,y).
\e
For $\hat{V}(x)=0$, we denote the solution to Eq.~(\ref{T}) as
\be
{\cal T}_x^{(0)}=\exp\lt[{\lt(\sigma_x\otimes \hat{q} +i\ep  \sigma_z\otimes \hat{\openone}\rt)x}\rt].
\e
The matrix ${\cal T}_L^{(0)}$ gives rise to the conductance 
of the ballistic strip of graphene, which was calculated in Ref.~\cite{Two06}, 
\be
\label{G0}
G^{(0)}=g_0 \s_n \lt[1 +\frac{q_n^2}{q_n^2-\ep^2}\sinh^2 \lt(L\sqrt{q_n^2-\ep^2}\rt)\rt]^{-1}.
\e
The Fermi-energy $\ep$ in the graphene sample 
is a monotonic function of the doping potential. The zero-temperature conductance (\ref{G0}),
plotted in Fig.~\ref{fig:doping} with the solid line, is minimal at $\ep=0$ and 
corresponds to $\sigma = 4 e^2/\pi h$. 
The minimal conductivity of ballistic graphene is due to the evanescent modes, 
which exponentially decay in the transport direction with rates $q_n$.

It is instructive to start with a simple impurity potential,
which is localized along a line $x=x_0$, 
\be
\label{V}
V(x,y)=\alpha(y) \delta(x-x_0).
\e
In this case the transfer matrix of the sample reads
\be
\label{T0a}
{\cal T}=
{\cal T}_{L-x_0}^{(0)} 
e^{-i\sigma_z\otimes\hat{\alpha}}
{\cal T}_{x_0}^{(0)}.
\e
At the Dirac point, $\ep=0$, we find from Eq.~(\ref{param}) 
\beq
\hat{t}^{-1} &=& \cosh (\hat{q} x_0) e^{i \hat{\alpha}} \cosh (\hat{q} (L-x_0)) \n \\
&+& \sinh (\hat{q} x_0) e^{-i \hat{\alpha}} \sinh (\hat{q} (L-x_0)),
\label{t1}
\eq
where the matrix elements of $\hat{\alpha}$ are given by the Fourier transform (\ref{FT}).
It is evident from Eq.~(\ref{t1}) that the conductance at $\ep=0$ is not 
affected by any potential located at the edges of the sample $x_0=0$ or $x_0=L$.

In order to maximize the effect of the impurity we let $x_0=L/2$
and calculate the conductance from Eqs.~(\ref{G},\ref{param},\ref{T0a}). 
We consider in detail two limiting cases for the $y$-dependence of $\alpha(y)$: 
a constant and a delta-function. 

For the constant potential,
\be
\label{const}
\alpha(y)=\alpha, 
\e
we have $\hat{\alpha}=\alpha \hat{\openone}$, hence we find, at $\ep=0$,
\be
\label{G00}
G=g_0\s_n\frac{1}{\cos^2\alpha\,\cosh^2 q_n L +\sin^2\alpha}.
\e
Note that for any $\alpha$ the conductance at the Dirac point is equal, 
or exceeds its value for $\alpha=0$. Moreover, the conductance is enhanced 
to $G=g_0 N$ if the parameter $\alpha$ equals one of the special values 
$\alpha_n\equiv \pi(n+1/2)$, where $n$ is an integer number. This is a resonant 
enhancement, which takes place only in a close vicinity of $\ep=0$.
Taking the limit $N\to \infty$ first, we obtain the logarithmic singularity
at the Dirac point $G=g_0 (W/\pi L) \ln |\ep|+ {\cal O}(1)$. The energy 
dependence of $G$ for the resonant values, $\alpha=\alpha_n$, 
is shown in Fig.~\ref{fig:doping} with the dashed line.

For a stochastic model with a fluctuating parameter $\alpha$,  
one finds a moderate enhancement of 
the averaged conductance due to the contribution of resonant configurations
with $\alpha \approx \alpha_n$. If the fluctuations of $\alpha$ have a large amplitude
(strong disorder), the contributions from different resonances are summed up 
leading to a universal result. In this case one can regard $\alpha$ as a random quantity, 
which is uniformly distributed in the interval $(0,2\pi)$. The averaged conductance 
for this model is plotted in Fig.~\ref{fig:doping} with the dotted line. 
It acquires the minimal value $G=g_0(W/2L)$ ($\sigma =2e^2/h$) at the Dirac point.
Note that the averaged conductance is enhanced as compared to that of a 
ballistic sample for $\ep < L /\hbar v$. For large doping 
the situation is opposite, i.e.~the conductance is suppressed by the impurity potential.

\begin{figure}[tb]
\centerline{\includegraphics[width=0.9\linewidth]{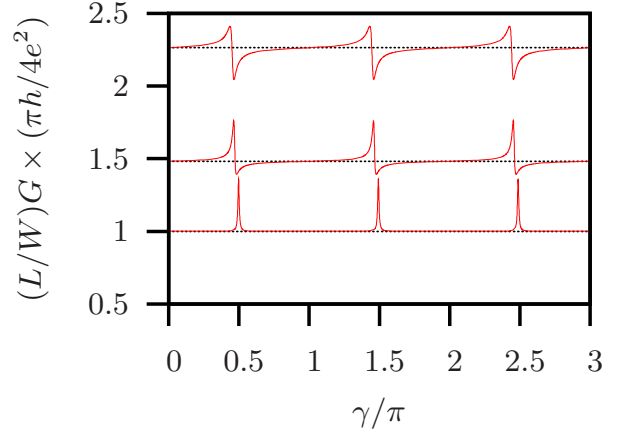}}
\caption{\label{fig:impurity} 
The conductance of the graphene strip, $W/L=7$, 
calculated from Eqs.~(\ref{G},\ref{param},\ref{T0a}) 
in the presence of a single delta-functional impurity 
(cf.~Eqs.~(\ref{V},\ref{delta}))
as a function of the impurity strength.
Different curves correspond to the different values of 
the chemical potential (Fermi energy) in the strip:  
$\ep L/\hbar v = 0$ for the lowest curve, $\ep L/\hbar v = 2$ for 
the curve in the middle, $\ep L/\hbar v=3$ for the upper curve. 
The dotted lines are guides to the eye.
}
\end{figure}

For the delta-function potential,
\be
\label{delta}
\alpha(y)=(\gamma W/M)\, \delta(y-y_0),
\e
we find the elements of the matrix $\hat{\alpha}$ 
in Eq.~(\ref{T0a}) as 
\be
\alpha_{nm}=(\gamma/M)e^{i(q_n-q_m)y_0}.
\e
Note that the ratio $W/M= \lambda_F$ remains constant in the limit $N\to \infty$.
We calculate the conductance from Eqs.~(\ref{G},\ref{param},\ref{T0a})
and plot the results in Fig.~\ref{fig:impurity} as a function of the parameter 
$\gamma$ for three different values of $\ep$. Since we assume periodic boundary 
conditions the conductance does not depend on the value of $y_0$. 

We see that the effect of a single delta-functional impurity is much smaller than
that of the constant potential (\ref{const}), but the main features remain.
At zero doping, $\ep=0$, the conductance is enhanced for the special values 
of the impurity strength $\gamma=\alpha_n$. Unlike in the previous case the 
height of the peaks is finite and is determined by the ratio $W/L$. (The effect 
is bigger if the impurity potential has a finite width.) It is clear from the lowest 
curve in Fig.~\ref{fig:impurity} that in the stochastic model of fluctuating $\gamma$ 
the conductance at the Dirac point is necessarily enhanced.

Away from the Dirac point the effect of the impurity is modified.
The conductance $G$ becomes an oscillating function of $\gamma$, 
and, for $\ep L \gg 1$, its value at finite $\gamma$ is 
smaller or equal the value at $\gamma=0$.

We have to stress that the conductivity enhancement at the Dirac point 
induced by disorder potential relies upon the symplectic symmetry 
of the transfer matrix ${\cal T}_x^T(-q) \sigma_x {\cal T}_x(q)=\sigma_x$,
which holds as far as the potential $V(x,y)$ is a scalar in the isospin space. 
The corresponding microscopic potential is smooth on atomic scales 
even in the limit $V(x,y)\propto \delta(\bb{r}-\bb{r}_0)$.

Let us illustrate the approach to disordered graphene, which follows from Eq.~(\ref{T}). 
The flux conservation justifies the parameterization
\be
\label{par}
{\cal T}_x= 
\lt(\ba{cc} \hat{U}_1 & 0 \\ 0 & \hat{U}^\prime_1\ea\rt)
\lt(\ba{cc} \cosh \hat{\lambda} & \sinh\hat{\lambda}\\
\sinh\hat{\lambda} & \cosh \hat{\lambda} \ea\rt)
\lt(\ba{cc} \hat{U}_2 & 0\\ 0 & \hat{U}^\prime_2 \ea\rt), 
\e
where $\hat{U}_{1,2}, \hat{U}^\prime_{1,2}$ are some unitary matrices in the channel space
and $\hat{\lambda}$ is a diagonal matrix.
The values $\lambda_n$ for $x=L$ determine the conductance of the 
graphene strip
\be
\label{Gl}
G=g_0\s_n \frac{1}{\cosh^2\lambda_n}.
\e
The detailed analysis of Eq.~(\ref{T}) in the parameterization (\ref{par})
is a complex task, which is beyond the scope of the present study.
The problem is greatly simplified in the ``one-dimensional'' limit
$V(x,y)=V(x)$ due to the absence of mode mixing. In this case 
the unitary matrices in the decomposition (\ref{par}) 
are diagonal and the prime corresponds to the complex conjugation.
We parameterize $\hat{U}_1=\diag\lt\{\exp{(i\theta_n)}\rt\}$, $n=-M,\dots M$,
and reduce Eq.~(\ref{T}) to a pair of coupled equations for each mode  
\beq
\label{1}
\frac{\pa \lambda_n}{\pa x} &=&q_n \cos 2\theta_n,\\
\frac{\pa \theta_n}{\pa x}&=& \ep-V(x)-q_n \sin 2\theta_n\,\coth 2\lambda_n.
\label{2}
\eq
For $V=\ep=0$, the transfer matrix fulfills an additional chiral symmetry,
hence $\theta_n=0$. In this case the variables $\lambda_n$ grow with 
the maximal rate $\lambda_n/x=q_n$ as $x$ increases, which corresponds to the minimal conductance.
Any finite doping, $\ep\neq 0$, or arbitrary potential $V(x)$ violates the chiral symmetry
and move the phases $\theta_n$ away from $\theta_n=0$. It follows from Eq.~(\ref{1})
that $\lambda_n < q_n x$, hence the conductance defined by Eq.~(\ref{Gl}) is enhanced
above its value for $V=\ep=0$. In general case of arbitrary $V(x,y)$ 
the conductance is enhanced only on average, since a rare fluctuations 
with suppressed conductance become possible. One illustration for the enhancement of the conductance 
in the presence of mode mixing is provided by the lowest curve in Fig.~\ref{fig:impurity}.

The effects of individual impurities on the resistivity of graphene samples 
in strong magnetic fields have been demonstrated in recent experiments \cite{Hwa06,Sch06}.
We, therefore, believe that the phenomenon of the impurity-assisted tunneling  
considered above allows for an experimental test.

Let us now give a brief analysis of the conductivity in the model with 
a one-dimensional disorder, which is described by a white-noise 
correlator in the transport direction  
\be
\label{pot}
\la V(x)V(x') \ra =\frac{1}{2\ell} \delta(x-x'),\quad \la V(x)\ra =0,
\e
and is assumed to be constant in the transversal direction. Even though such 
a choice of disorder potential is clearly artificial, it gives rise to an 
an analytically tractable model. Due to the absence of mode mixing 
we can omit the index $n$ in Eqs.~(\ref{1},\ref{2}) and study the 
fluctuating variable $\lambda$ as a function of $q$ and $L$. 
The two-terminal conductivity $\sigma=L G/W$, where $G$ is found from Eq.~(\ref{Gl}),
is given in the limit $W\to \infty$ by the integral over the transversal momentum 
\be
\label{sig}
\sigma=g_0\, L \int_{-\infty}^\infty \frac{dq}{2\pi} \frac{1}{\cosh^2 \lambda(q,L)}.
\e

We note that Eqs.~(\ref{1},\ref{2}) with the white noise potential 
(\ref{pot}) are analogous to the corresponding equations  
arising in the problem of Anderson localization on a one-dimensional lattice 
in a vicinity of the band center \cite{Sch03}. 

We look for the solution in the limit of large system size $L\gg \ell$ 
in which case the standard arguments can be applied. 
First of all, the variable $\lambda$ is self-averaging in the limit $L\gg \ell$, 
therefore the mean conductivity can be estimated by the substitution 
of the averaged value of $\lambda$ in Eq.~(\ref{sig}),
\be
\la \lambda \ra = \int_0^L dx\, \la \cos2\theta \ra  
\simeq q L \la \cos 2\theta \ra,
\e
where the mean value of $\cos 2\theta$ in the last expression is to be 
found from the stationary probability density $P(\theta)$ of the phase variable.
The main contribution to the integral in Eq.~(\ref{sig}) comes from $\lambda \sim 1$ 
since very small values of $\lambda$ are not affected by disorder. As the result 
we can let $\coth \lambda \sim 1$ in Eq.~(\ref{2}) and derive the 
Fokker-Planck equation on $P(\theta)$ in the stationary limit $L\gg \ell$ 
\be
\label{FP}
\ep\frac{\pa P}{\pa \theta}+q\frac{\pa}{\pa\theta}\sin 2\theta P+
\frac{1}{4 \ell} \frac{\pa^2 P}{\pa \theta^2}=0.
\e
The solution to Eq.~(\ref{FP}) has the form
\be
P(\theta)\propto \int_0^\infty \!\!dt\,\, \exp\lt[-4 \ep \ell t+4 q\ell \sin t \,\sin (t-2\theta)\rt],
\e
which leads to 
\be
\label{I}
\la \lambda(q,L) \ra = qL \frac{\int_0^\infty dt\, e^{-4 \ep \ell t}I_1(4 q \ell \sin t) \sin t}
{\int_0^\infty dt\, e^{-4 \ep \ell t}I_0(4 q \ell \sin t)},
\e
where $I_0$, $I_1$ stay for the Bessel functions.

We notice that in the limit $L\gg \ell$ the integral in Eq.~(\ref{sig}) 
is determined by the modes with $q\ell \ll 1$. For such modes we can let
$I_1(u)=u/2$, $I_0(u)=1$ in Eq.~(\ref{I}) and obtain 
\be
\label{lam}
\la \lambda \ra = q^2 \ell L\,\frac{1}{1+(2\ep\ell)^2}.
\e
An interesting observation can be made at this stage. 
Exploiting the analogy with Anderson localization a bit further 
we introduce a notion of the mode-dependent localization length $\xi$ 
from the relation $\la \lambda \ra = L/\xi$. We, then, arrive at the standard
result $\xi = (\ep/q)^2 \ell$ (which means that the localization length 
is set up by the mean free path) only in the limit of large doping 
$\ep\ell \gg 1$. On contrary, for $\ep \ell \ll 1$ we find the counterintuitive 
inverse dependence $\xi = (q^2 \ell)^{-1}$. 
This emphasizes once again an intimate relation of the underlying physics 
to the disorder-assisted tunneling \cite{Fre96,Luc04}, 
which indeed suggests an enhancement of the length $\xi$ with increasing disorder strength. 

Substitution of Eq.~(\ref{lam}) to Eq.~(\ref{sig}) yields
\be
\label{sigres}
\sigma = c\,g_0 \sqrt{\frac{L}{\ell}}\sqrt{1+\lt(\frac{2\ep\ell}{\hbar v}\rt)^2}, \quad L\gg \ell,
\e
with the constant $c \approx 0.303$. Thus, the two-terminal conductivity
in the model with one-dimensional fluctuations of the disorder potential 
increases with the system size without a saturation.
The width of the conductivity minimum is essentially broadened 
by disorder and is defined by the inverse mean free path $\hbar v/\ell$ 
instead of the inverse system size $\hbar v/L$ in the ballistic case.
In the calculation presented above we have chosen to average 
$\lambda$ rather than $\cosh^{-2}\lambda$. This cannot affect 
the functional form of the result (\ref{sigres}), however, 
the numerical constant $c$ can slightly depend on 
the averaging procedure. 

Even though the localization effects are very important in the derivation of 
Eq.~(\ref{sigres}), the Anderson localization in its original sense is absent 
in the considered model. Indeed, Eq.~(\ref{sigres}) assumes that the conductance 
of the graphene strip decays as $L^{-1/2}$. 

A generic random scalar potential in Eq.~(\ref{H}) would lead to the mode-mixing 
unlike the specific random potential (\ref{pot}) considered above. 
It is natural to expect, on the basis of the single-impurity analysis (\ref{delta}), 
that the mode-mixing will strongly suppress the effect of the conductivity 
enhancement. Nevertheless, the size dependent growth of the conductivity 
has been conjectured in Ref.~\cite{Lud94} for a model of Dirac fermions 
in a generic two-dimensional scalar disordered potential. 
Moreover, the very recent numerical studies \cite{Bar07} provide a solid evidence
of the logarithmic increase of the conductivity with the system size. 
The behaviour observed in \cite{Bar07} can be described 
by the expression of the type (\ref{sigres}) provided $\sqrt{L/\ell}$ 
is replaced by $\ln (L/\ell)$, where $\ell$ is inversely proportional 
to the potential strength. These results are in sharp contradiction 
with the recent work by Ostrovsky {\it et al.} \cite{Ost07}, where a non-trivial 
renormalization-group flow with a novel fixed point corresponding 
to a scale-invariant conductivity $\sigma^* \approx 0.6\, g_0$ 
is predicted. 

The conductivity enhancement discussed above relies upon the symplectic 
symmetry of the model (\ref{H}). We should remind that quantum interference effects 
are also responsible for a size-dependence of conductivity of a disordered normal metal. 
In two dimensions, the zero temperature conductivity, 
which includes the weak-localization correction, takes the well-known form 
\be
\label{WL}
\sigma_{{\rm 2D\, metal}} =
\frac{n e^2 \tau_s}{m} \pm \frac{2 e^2}{\pi h} \ln \frac{L}{\ell},
\e
where $n$ is the density of states at the Fermi level, $\tau_s$ is the
scattering time, and $\ell$ is the mean free path. The positive sign 
in Eq.~(\ref{WL}) corresponds to the case of a strong spin-orbit 
scattering \cite{Hik80} (the symplectic symmetry class). Thus, in a normal metal 
the symplectic symmetry also gives rise to the conductivity enhancement. This is 
in contrast to the orthogonal symmetry, which leads to 
the negative correction in Eq.~(\ref{WL}).
The derivation of Eq.~(\ref{WL}) takes advantage of the 
small parameter $(k_F \ell)^{-1}$ and cannot be generalized 
to graphene at low doping. Nevertheless the numerical results 
of Ref.~\cite{Bar07} can be formally 
described by applying Eq.~(\ref{WL}) beyond its validity range; 
i.e. in the situation when the Drude contribution 
(given by the first term in Eq.~(\ref{WL})) is disregarded 
in the vicinity of the Dirac point as compared to the weak localization term. 
We note, however, that the weak localization is not the only source 
of the logarithmic size dependence of conductivity in graphene \cite{Ale06}. 

In summary, an impurity potential, which is smooth on atomic scales,
improves the conductance of undoped graphene. A confined potential
can lead to a greater enhancement of the conductance 
than the uniform doping potential. One single impurity 
can noticeably affect the conductance   
provided its strength is tuned to one of the multiple resonant values.
We develop the transfer-matrix approach to the disordered graphene
and calculate the two-terminal conductivity in the model 
of one-dimensional potential fluctuations. The resulting 
conductivity is fully determined by interference effects and 
increases as the square root of the system size.

I am thankful to J.~H.~Bardarson, C.~W.~J.~Beenakker, P.~W.~Brouwer, 
and J.~Tworzyd{\l}o for numerous discussions and
for sharing with me the results of Ref.~\cite{Bar07} 
before publication. Discussions with W.~Belzig, J.~Lau, and M.~M\"uller 
are gratefully acknowledged. 
This research was supported in part 
by the German Science Foundation DFG through SFB 513.

\end{document}